# Marketplaces for Energy Demand-Side Management based on Future-Internet Technology

Luigi Briguglio[1], Frank Eichinger[2], Massimiliano Nigrelli[1], Javier Lucio Ruiz-Andino[3]

[1] Engineering Ingegneria Informatica S.p.A., Rome, Italy
`luigi.briguglio@eng.it, massimiliano.nigrelli@eng.it`
[2] SAP AG, Karlsruhe, Germany
`f.eichinger@sap.com`
[3] Telefonica Investigación y Desarrollo, Madrid, Spain
`lucio@tid.es`

**Abstract.** Renewable energies become more important, and they contribute to the EU's goals for greenhouse-gas reduction. However, their fluctuating nature calls for demand-side-management techniques, which balance energy generation and consumption. Such techniques are currently not broadly deployed. This paper describes the latest results from the FINSENY project on how Future-Internet enablers and market mechanisms can be used to realise such systems.

## 1  Introduction

Renewable energies help to reduce greenhouse gas emissions and thus to contribute to the goal of the European Union to reduce these emissions by 80% [1]. However, renewable energies also become more and more a challenge for our electricity grids. In addition, electric mobility can cause that distribution grids reach their limits. Especially the charging stations are new unpredictable heavy loads. Electricity grids need to be stable and must allow everybody at any point in time to consume energy. This requires that electricity grids have to permanently maintain a balance between demand and supply. However, renewable energy sources and unpredictable loads refer to an unsteady and fluctuating production and consumption of energy. Particularly the production of photovoltaic or wind energy does not necessarily match the energy demand patterns of consumers, but depends on weather conditions. Compensation for the fluctuating nature of renewables can be done, e.g., with flexible gas turbines in stand-by operation. However, this is very expensive. An alternative to additional production is to shift demands. Respective programmes are frequently referred to as *demand-side management (DSM)* or *demand response (DR)*, and electricity grids featuring such techniques are often called *smart grids*. In the scientific literature, a wide range of DSM approaches building on control theory and market mechanisms have been suggested to balance renewable generation and consumption (see, e.g., [2–5]).
A further motivation for the need of shifting energy demand is the distributed nature of renewable energy sources. This challenges the electricity system, which has been designed based on the understanding and technology that power flows from higher voltage levels (from centralised power plants) to lower voltage levels (to the consumers). Distributed generation units however typically operate at the lower voltage levels, which can potentially result in a unidirectional power flow for which the electricity grids have not been build. DSM can help to realise a paradigm shift from demand-driven generation in the past to generation-driven demand in the future. This would help to deal with both the fluctuating and distributed nature of renewable energies.

Despite the strong need for DSM solutions to facilitate a renewable energy supply and new loads from electric mobility, DSM solutions are in most countries not broadly available (apart from some energy-intensive industrial consumers). This is partly due to low incentives and missing legal frameworks, but it is also due to missing ICT infrastructure including scalability, security and interoperability aspects. Future-Internet technology promises to reduce time-to-market and to facilitate the deployment of new services by providing a unified platform. This helps to decrease operational costs by handling common tasks such as management of devices, user authentication and interchange of information between different actors, domains, devices and systems. We therefore build our DSM architecture on Future-Internet technology.

While marketplaces are commonly used to trade electrical energy [6], marketplaces for DSM have rarely been employed outside of research projects such as MeRegio [7] and NOBEL [8]. In this paper, we propose a marketplace-based architecture for DSM, which is the focus of our current research activities in the FINSENY project [9]. It builds on a common Future-Internet ICT architecture. In particular, we build this architecture on generic and domain-specific enablers stemming from the FI-WARE project [10]. This architecture facilitates the communication and participation in DSM programmes. We expect that the use of Future-Internet enablers will make it dramatically easier to realise our marketplace scenario in real-world implementations. This will facilitate a broader penetration of DSM technology and the integration of renewable energies and electric mobility with the electricity grids.

## 2    Overview of the Marketplace-based DSM Solution

When energy retailers experience energy shortages during a day, they buy energy at intra-day power exchanges [6]. It might however be cheaper to ask their customers via DSM to consume less energy within a certain time frame. Similarly, grid operators monitor the electricity grid and may need DSM to temporarily reduce their consumption in order to achieve grid stability. Besides the reduction of consumption, DSM mechanisms can equally be used to increase consumption in a certain time frame, e.g., when renewable production is particularly high (in certain grid segments).

This paper describes a marketplace-based approach for DSM. It has a focus on solving grid issues (i.e., consumption needs to be decreased temporarily), but it is equally suited for decreases and increases to optimise procurement of energy retailers. In particular, our approach is based on the idea of having two marketplaces:

1. a *B2C marketplace* where customers can subscribe to different DSM programmes,
2. a *B2B marketplace* where so-called *demand-side managers* (*DSM Managers* in Figure 1) can trade their bids for shifting energy demand.

The role of the demand-side manager may be taken on by the local distribution system operator, by energy retailers or by dedicated demand-side-manager companies. In this paper, we assume an infrastructure where customers have installed an *energy-efficiency control system (EECS)* in their premises that monitors and controls the energy consumption of appliances, by changing their programming parameters. This system allows receiving DSM signals from the demand-side manager based on the subscribed conditions and user preferences.

This architecture with two marketplaces is different from existing approaches. In traditional energy marketplaces such as power exchanges, electrical energy is traded as such [6]. Our B2B marketplace however trades energy shifts, which is not commonly done outside of research projects. As one example, the NOBEL project trials a marketplace for neighbourhoods where prosumers (consumers who also produce energy) can deal both energy and energy shifts [8]. Our architecture entirely relies on Future-Internet technology and integrates the internal market of energy utilities (B2B marketplace) with the market facing real customers (B2C marketplace).

**The B2C Marketplace**

Customers have access to the B2C marketplace on the Internet where they can see different offerings for DSM programmes from demand-side managers. These offerings could be based on real-time tariff schemes in case that the energy retailer is also the demand-side manager. Customers can then choose to contract one of these services. This aims at reducing the monthly bills. The offers might be coupled with energy contracts and vary in the incentives on how much money is paid (or reduced from the bill) when a customer actually reacts to DSM signals. By contracting such a service, the customer allows the demand-side manager to send DSM signals to her or his EECS. These signals are used to initiate actions in the appliances connected to the EECS, in order to schedule operations (e.g. electric-vehicle charging, starting a washing machine, anticipating/delaying a production pipeline, lowering/increasing the temperature of the customers building by some degrees within a certain range). It must be noted that the contract needs to reflect the possibility for the customer to not accept certain DSM signals at any time in order to allow customer's manual decisions on how and when to use the contracted energy.

**The B2B Marketplace**

When customers have contracted with demand-side managers and an infrastructure is available that can execute DSM measures, the B2B marketplace described in the following uses market mechanisms for trading flexible loads for DSM purposes. Energy retailers considering a DSM measure can send their request to the electronic B2B marketplace (Step 1 in the architecture overview in Figure 1). Equally, if overloads or voltage problems are detected, the grid operator can issue a similar request (Step 1). Then, all affected demand-side managers receive these requests from the B2B marketplace (Step 2), and each demand-side manager can place an offer (a bid) including a price for reacting to the request (Step 3). The B2B marketplace then selects a com-bination of offers which fulfils the request based on economic principles (Step 4). If the retailer does not accept the assembly of offers, the retailer might prefer to alternatively (Step 5a) buy energy at the exchange if this is cheaper (Step 6a). If the retailer or grid operator accepts the assembly of offers, the respective demand-side managers are then responsible to conduct the DSM request (Step 5b/6b). To do so, the demand-side managers then send DSM signals to the EECSs of their contracted customers (Step 7). These systems then take the best decision based on the signals, incentives, available electricity (e.g., from photovoltaic generation), user preferences and consumption data to send the signal to intelligent devices in the customer's premises as well as charging infrastructure of electric vehicles (Steps 8–10). However, the customer is able at any moment to avoid DSM signals using a dedicated EECS graphical user interface (GUI; Step 9a).

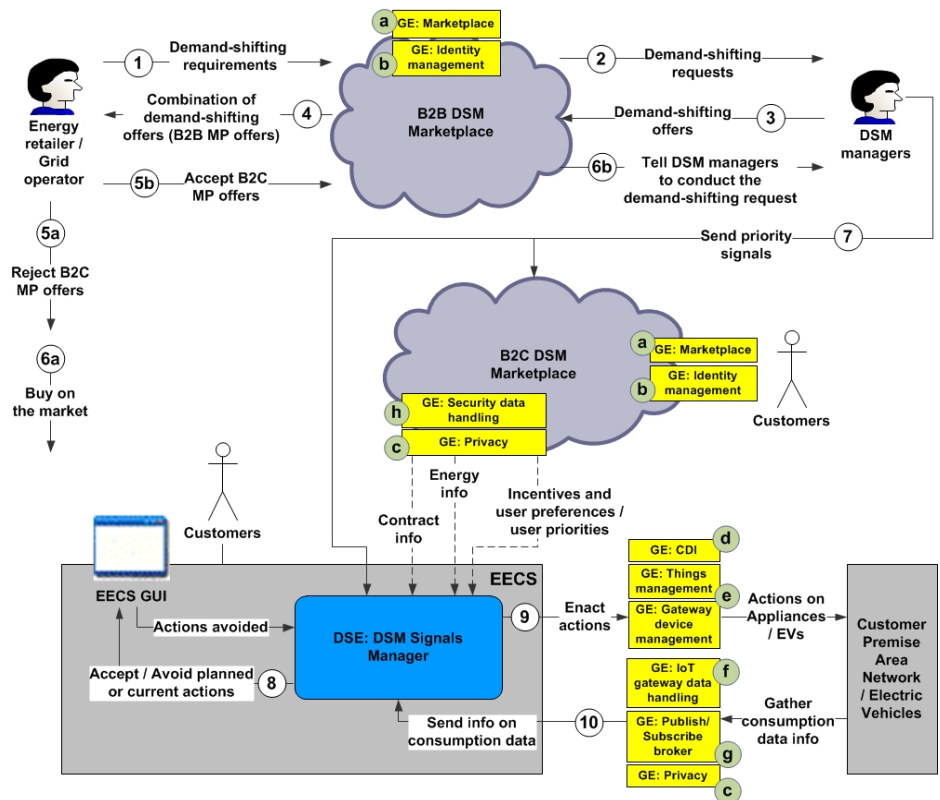

**Figure 1: Architecture Diagram**

## 3   DSM Marketplaces based on Future-Internet Technology

Future-Internet technologies introduce innovative infrastructures for cost-effective creation and delivery of services, providing high QoS and security guarantees. Its use in the DSM solution proposed will reduce the time-to-market and accelerate business opportunities, will facilitate the deployment of new services by providing a complete platform, and will decrease operational costs by providing common operation tools of the DSM marketplaces. Moreover, they will allow the interchange of information across the different marketplace actors, domains, devices and systems, and will provide a standardised way for the respective communications and for the configuration and management of the devices. In a more concrete way, Future-Internet technologies will offer several opportunities for smart-energy solutions. These include: (i) Connectivity between the different actors and grid elements such as DSM managers, EECSs, renewable energy resources, charging stations, final customers etc. using common and public communication infrastructures (2G/3G networks, LTE, IPv6, etc.) and allowing real-time communications needed for an efficient management of the DSM marketplaces, network virtualisation techniques etc. (ii) Management of the increased data volume, complexity and cost due to the DSM requirements as registry/repository of actors, devices, data mediation, billing, contracts, business models, dynamic energy prices etc. (iii) Service enablement allowing new ways of collaboration between DSM marketplace actors by providing new web services based on bi-directional communication and interaction enabled by advanced ICT solutions. (iv) Decentralisation and distribution of the intelligence of the grid enabling the participation of every DSM actor for information, coordination and control the relationship between them. Telecommunication networks, as the Internet, have demonstrated that using advanced ICT solutions is essential to achieve this objective. (v) Security and privacy support that have a very important role in the DSM operation for guaranteeing the highest security standards and privacy of personal data in the marketplace.

**Use of Generic and Specific Future-Internet Enablers**

The architecture proposed in this paper is based on the identification of generic enablers (GE) from the FI-WARE project [11] and in a specific domain specific enabler (DSE) for the management of the DSM signals. Figure 1 illustrates this architecture. In particular, several generic enablers are involved in the DSM marketplace: (a) Marketplace: managing contracts, generating reports or automated reminders, providing the functionalities to offer and deal with services and to combine them to value-added services for customers. (b) Identity management: providing authentication/access control and identity/attribute assertions as a service to relying parties. (c) Privacy: providing a set of functionality for privacy enhancing technologies. (d) Connected devices interfacing (CDI): detecting and optimally exploiting capabilities and aspects about the status of devices involved in the DSM marketplace. (e) Gateway devices and things management: independent from the technology involved in each device. (f) IoT gateway data handling: collecting large amounts of DSM-related data produced by the DSM actors in real-time. (g) Publish/subscribe broker: publication of related DSM events by the different actors so that they become available to other DSM actors. (h) Security data handling: providing the security policies for accessing data.

## 4    Conclusions

Renewable energies and electric mobility challenge the electricity grids. Demand-side management (DSM) can help to solve this, but introduces several ICT requirements. Future-Internet technologies offer various opportunities for dealing with these requirements and to facilitate a broader penetration of DSM. Future-Internet technologies allow, among others, the interchange of information across multiple domains and stakeholders, among devices and between subsystems of diverse complexity in a standardised way. Moreover, these technologies will ensure the interoperability and correct configuration of these actors and management of their boundaries.

The results partially presented in this paper will be refined in the project FINSENY. A following project (FINESCE) will investigate and evaluate in a trial how the concepts described in this paper could practically solve grid issues in the city of Terni, Italy.

## Acknowledgements

The authors gratefully acknowledge the contributions from Pasquale Andriani, Valter Bella, Giuseppa Caruso and Pierre Plaza Tron. Several concepts in this paper regarding the B2B marketplace are inspired by the project MeRegio [7].

This research has received funding from the Seventh Framework Programme of the European Union (FP7/2007–2013) under grant agreement no. 285135 (FINSENY – Future INternet for Smart ENergY [9]; http://www.fi-ppp-finseny.eu/).